\documentclass[twocolumn, prl]{revtex4-1}
\usepackage{amssymb}
\usepackage{amsmath}
\usepackage{epsfig}
\usepackage{color}
\usepackage{graphics, graphicx}
\usepackage{bbold}
\usepackage{psfrag}
\usepackage{mathcomp}
\usepackage{subfigure}
\usepackage{verbatim}
\usepackage{color}
\usepackage[colorlinks, citecolor=blue]{hyperref}

\begin{document}
\title{Resonant Scattering and Microscopic Model of Spinless Fermi Gases in One-dimensional Optical Lattices}
\author{Xiaoling Cui}
%\email{xlcui@iphy.ac.cn}
\affiliation{Beijing National Laboratory for Condensed Matter Physics, Institute of Physics, Chinese Academy of Sciences, Beijing 100190, China}
\date{\today}
\begin{abstract}
We study the effective Bloch-wave scattering of a spinless Fermi gas in one-dimensional (1D) optical lattices. By tuning the odd-wave scattering length, we find multiple resonances of Bloch-waves scattering at the bottom (and the top) of the lowest band, beyond which an attractive (and a repulsive) two-body bound state starts to emerge. These resonances exhibit comparable widths in the deep lattice limit, and the finite interaction range plays an essential role in determining their locations.  Based on the exact two-body solutions, we construct an effective microscopic model for the low-energy scattering of fermions. The model can reproduce not only the scattering amplitudes of Bloch-waves at the lowest band bottom/top, but also the attractive/repulsive bound states within a reasonably large energy range below/above the band. These results lay the foundation for quantum simulating topological states in cold Fermi gases confined in 1D optical lattices. 
\end{abstract}

\maketitle

{\it Introduction.} As a prominent example of quantum simulation, an odd-wave interacting Fermi gas in one-dimensional(1D) optical lattices can serve as an ideal platform for realizing the Kitaev chain model\cite{Kitaev}, a prototype of system hosting Majarona fermions\cite{Majarona} and recently attracting great attention in condensed matter physics\cite{Majarona_solid}.  
Nevertheless, to achieve the goal of quantum simulation using cold atoms, %accurately simulate  intriguing many-body physics in optical lattices, 
it is fundamentally important to understand the two-body scattering property of a dilute gas as the first step. For instance, it has been found that the interplay between strong s-wave interaction and lattice potential can significantly modify the low-energy scattering property of Bloch waves%, and even induce Bloch-wave resonances and bound state
\cite{Zoller, Ho, Buchler, Cui, Stecher, Carr}. Lattices can also support a peculiar type of repulsive bound state that is excluded in continuum\cite{rep_bound}. Furthermore, the correct understanding of two-body scattering property is the foundation to construct an effective low-energy model, which will facilitate the study of many-body physics as the next step.

In this work, we exactly solve the two-body effective scattering of odd-wave interacting (spinless) fermions in 1D optical lattices. 
We adopt a two-channel Hamiltonian that naturally incorporates the effect of finite interaction range, as a realistic situation in cold atoms when reducing the 3D p-wave interacting Fermi gas\cite{K40, K40_2, Li6_1, Li6_2} to quasi-1D by transverse confinement\cite{CIR_p_expe, CIR_p_1, CIR_p_2, CIR_p_3}. %, as a realistic situation in current cold atoms experiments. 
Based on the recently developed interaction renormalization approach for 1D odd-wave  systems\cite{Cui2, Cui3}, our formulism are able to capture all the high-band effects and applicable to arbitrary lattice depths and interaction strengths/ranges. The main findings include (i) the multiple Bloch-wave resonances by tuning odd-wave scattering strengths and associated attractive/repulsive bound states; (ii) the sensitive dependence of resonance locations and widths on the interaction range and the lattice depth; (iii) an effective model constructed for lowest-band fermions, which correctly predicts both the scattering amplitudes of Bloch waves and the bound states below/above the lowest band. These results reveal the unique scattering property due to the interplay of odd-wave interactions and lattice potentials, and pave the way for future exploring the physics of Majorana fermions in cold atomic gases. % of interacting spin-less Fermi gases in optical lattices.

% Model, formula
{\it Formulism.} We start from a two-channel Hamiltonian:
%model of spin-less fermions in 1D lattices: of two spin-less fermions in the open channel and one dimer in the closed channel: 
\begin{eqnarray}
%{\cal H}=\int dx \left( \psi^{\dag} h_f \psi + d^{\dag} h_d d \frac{g}{2}\left( d^{\dag} [(i\partial\psi)\psi-\psi(i\partial\psi)] +h.c. \right)\right) \label{H0}
{\cal H}&=&\int dx \left( \psi^{\dag} h_f \psi + d^{\dag} h_d d +U_{fd}\right); \label{H0}\\
U_{fd}&=&\frac{g}{2}\left( d^{\dag} [(i\partial\psi)\psi-\psi(i\partial\psi)] +h.c.\right).\nonumber\end{eqnarray}
Here $\psi^{\dag}$ and $d^{\dag}$ are respectively the creation operators of open-channel fermions and closed-channel dimers under single-particle Hamiltonian $h_f=-\partial_x^2/(2m) + V_f(x)$ and  $h_d= -\partial_x^2/(4m)+\nu +V_d(x)$, with lattice potentials $V_f=V_0 \sin^2(\pi x/a_L)$ and $V_d=2V_f(x)$; $\nu$ is the closed-channel detuning, and $g$ is the coupling strength between two channels. The free-space scattering length $l_o$ and effective range $r_o$ for odd-wave interaction
%can be related to $g$ and $\nu$ 
are defined through renormalization equations\cite{Cui2,Cui3}:
%$m/(2l_o)=-\nu/(2g^2) + L^{-1} \sum_q q^2/(2\epsilon_q)$ and $r_o=(m^2g^2)^{-1}$ (here $\epsilon_q=q^2/(2m)$; 
\begin{eqnarray}
\frac{m}{2l_o}&=&-\frac{\nu}{2g^2} + \frac{1}{L} \sum_q \frac{q^2}{2\epsilon_q}; \label{lo}\\
r_o&=&\frac{1}{m^2g^2}; \label{ro}
\end{eqnarray}
where $\epsilon_q=q^2/(2m)$ and $L$ is the length of the system. Here we consider the p-wave resonance of identical $^{40}$K fermions near 200G\cite{K40,K40_2} under a tight transverse confinement with frequency $\omega_{\perp}$, and $\omega_{\perp}$ sets the largest energy scale in this paper so that the system is effectively in 1D regime. 
%In this paper all the energies are much smaller than $\omega_{\perp}$, so that the system is effectively in 1D regime.
%In this paper all the physical energies considered are much smaller than $\omega_{\perp}$, so that the system is effectively moving in 1D. }
Given $a_{\perp}=1/\sqrt{m\omega_{\perp}}\sim 50nm$ and a large 3D p-wave range $\sim4\times 10^{6}{\rm cm}^{-1}$, an estimation based on Ref.\cite{CIR_p_1, CIR_p_2, CIR_p_3} gives $r_o\sim 250$nm, which is about half of typical lattice spacing $a_L\sim 500$nm. Thus in this paper we take $r_o=0.5 a_L$ and use $k_L=\pi/a_L$ and $E_L=k_L^2/(2m)$ as the units of momentum and energy, respectively. 
In particular, we scale the lattice depth as $v\equiv V_0/E_L$. %($\hbar$ is set to be unity). 

Expanding $\psi,\psi^{\dag}$ and $d,d^{\dag}$ 
%the fermion and dimer field operators 
in terms of the Bloch wave eigenstates of $h_f$ and $h_d$, the Hamiltonian (\ref{H0}) can be rewritten as:
\begin{eqnarray}
{\cal H}&=&\sum_{nk} \epsilon_{nk} \psi^{\dag}_{nk}\psi_{nk} + \sum_{NK} (E_{NK} + \nu) d^{\dag}_{NK}d_{NK}   \nonumber\\
&&+ \frac{g}{\sqrt{L}} \sum_{NK} \sum_{nn';kk'}\left(  c_{nn';kk'}^{NK} d^{\dag}_{NK} \psi_{nk}\psi_{n'k'}+h.c. \right). \label{H1}
\end{eqnarray}
Here $\epsilon_{nk}$ and $E_{NK}$ are respectively the Bloch-wave energies of fermions and dimers, with 
%$\phi_{nk}(x)=\sum_G a_n(k+G)e^{i(k+G)x}/\sqrt{L}$ and $\Phi_{NK}(x)=\sum_Q A_N(K+Q)e^{i(K+Q)x}/\sqrt{L}$ 
%$\phi_{nk}(x)$ and $\Phi_{NK}(x)$ with eigen-energies $\epsilon_{nk}$ and $E_{NK}$,  where 
$n,N=\{0,1,2...\}$ the band index and $k,K\in(-k_L,k_L]$ the crystal momentum, and
$c_{nn';kk'}^{NK}$ is the atom-dimer coupling constant. %\cite{supple}. 
%and the coupling is 
%\begin{equation}
%%c_{nn'k}^{NK}= \sum_{Q,G} \left(K+G-\frac{K+Q}{2}\right) A_N^*(K+Q)a_n(k+G)a_{n'}(K+Q-k-G)
%c_{nn';kk'}^{NK}=\frac{i}{2} \int dx  \Phi_{NK}^*(x) \left(\frac{\partial \phi_{nk}}{\partial x}\phi_{n'k'}(x)- \frac{\partial \phi_{n'k'}}{\partial x}\phi_{nk}(x)\right).
%\end{equation}
One can check that a non-zero $c_{nn';kk'}^{NK}$ requires $k+k'$ identical to $K$ up to an integer number of $2k_L$. Therefore $K$ is a good number during the scattering process. 

We write down the two-body ansatz with given $K$:
\begin{equation}
|\Psi\rangle_K=\frac12 \sum_{nn'}\sum_{k} \alpha^{K}_{nn';k}\psi^{\dag}_{nk}\psi^{\dag}_{n',[K-k]} +\sum_N \beta^K_N d^{\dag}_{NK}, \label{psi}
\end{equation}
here $[..]$ is to shift $K-k$ by an integer number of $2k_L$ to be within $(-k_L,k_L]$. By imposing the Schr{\" o}dinger equation ${\cal H}|\Psi\rangle_K=E|\Psi\rangle_K$, we obtain the coupled equations:  %for $\{\alpha^K_{nn';k}\}$ and $\{\beta^K_N\}$:
\begin{eqnarray}
(E-\epsilon_{nk}-\epsilon_{n',[K-k]})\alpha^{K}_{nn';k} &=& 2g\sum_N c_{nn';k,[K-k]}^{NK\ *} \beta_N^K ;\\
(E-E_{NK}-\nu)\beta_N^K&=&g\sum_{nn';k} c_{nn';k,[K-k]}^{NK} \alpha^{K}_{nn';k}.
\end{eqnarray}
By eliminating $\alpha^K_{nn';k}$, these equations can be reduced to
\begin{equation}
\left( \frac{m}{2l_o} -M^K_{NN} \right) \beta^K_N=\sum_{N'\neq N} M^K_{NN'}\beta^K_{N'} \label{RG}
\end{equation}
with
\begin{eqnarray}
M^K_{NN'}&=&\left( -\frac{m^2r_o}{2}(E-E_{NK}) +\frac{1}{L}\sum_k \frac{k^2}{2\epsilon_k} \right) \delta_{NN'} \nonumber\\
&&+ \frac{1}{L}\sum_{nn';kk'} \frac{c_{nn';kk'}^{NK} c_{nn';kk'}^{N'K\ *} }{E-\epsilon_{nk}-\epsilon_{n'k'}}. \label{M}
\end{eqnarray}
Note that the relation $k'=[K-k]$ is hidden in above summation to ensure the finite $c_{nn';kk'}^{NK}$. In writing Eqs.(\ref{RG},\ref{M}), we have also utilized the renormalization equations (\ref{lo},\ref{ro})\cite{Cui2,Cui3}. We see that here the lattice affects the low-energy solution ($E$) through the modification of spectra ($\epsilon_{nk},\ E_{NK}$) and couplings ($c_{nn';kk'}^{NK}$). Since the lattice does not affect the scattering in high-energy space, the two ultraviolet divergences in $M_{NN}^K$ can exactly cancel with each other. 

We remark that Eq.\ref{RG} can apply to different  interaction strengths/ranges, lattice depths and total momenta for spinless fermions scattering in 1D lattices. % to any has incorporated all essential effects due to interactions and the lattice.
%, to the two-body problem. Its physical interpretation can be understood as follows. 
Its left and right sides respectively describe the scattering process within each dimer level and between different levels, while the latter is caused 
%, while the right the off-diagonal scattering between different dimer levels, 
by the coupling between relative and center-of-mass motions due to the presence of lattice potentials\cite{similar_s_wave}. %Eq.\ref{RG} works for any interaction parameters,  lattice depths and total momenta $K$. 
%Similar coupled equations also exist for s-wave interacting fermions in lattices\cite{Buchler, Carr}. 
Here we will focus on the two-body ground state with $K=0$, and simplify $c_{nn';kk'}^{N,K=0}$ as $c_{nn';k}^{N}$.

{\it Bound state spectrum.} The bound state solution $E=E_b$ can be obtained by requiring nontrivial solutions of $\{\beta_N^K\}$ in Eq.\ref{RG}. In Fig.1a, we show $E_b$ as a function of $1/l_o$ at given $r_o=0.5a_L$ and $v=6$. As increasing $1/l_o$, we can see a series of bound states emerging from the two-body continuum, corresponding to the coupled-channel ($\{N\}$) solutions in Eq.\ref{RG}. 
%correspond to the coupled-channel are found to distribute below the lowest scattering band, and between different band continuum. These bound states correspond to multiple dimer levels involved in the problem. 
Given the property that $M_{NN'}$  is finite only for even $N-N'$, these bound states  fall into two classes: one is by coupling dimer levels with even $N$(solid lines in Fig.1a), which produces an even-parity two-body wave-function in the center-of-mass motion: $\Psi(x_1,x_2)= \Psi(-x_2,-x_1)$; the other is by coupling all odd-$N$ levels (dashed lines) and produces an odd-parity wave function: $\Psi(x_1,x_2)=- \Psi(-x_2,-x_1)$. We see that the ground state belongs to the even-parity class.

\begin{figure}[h]
\includegraphics[width=8cm,height=5.5cm]{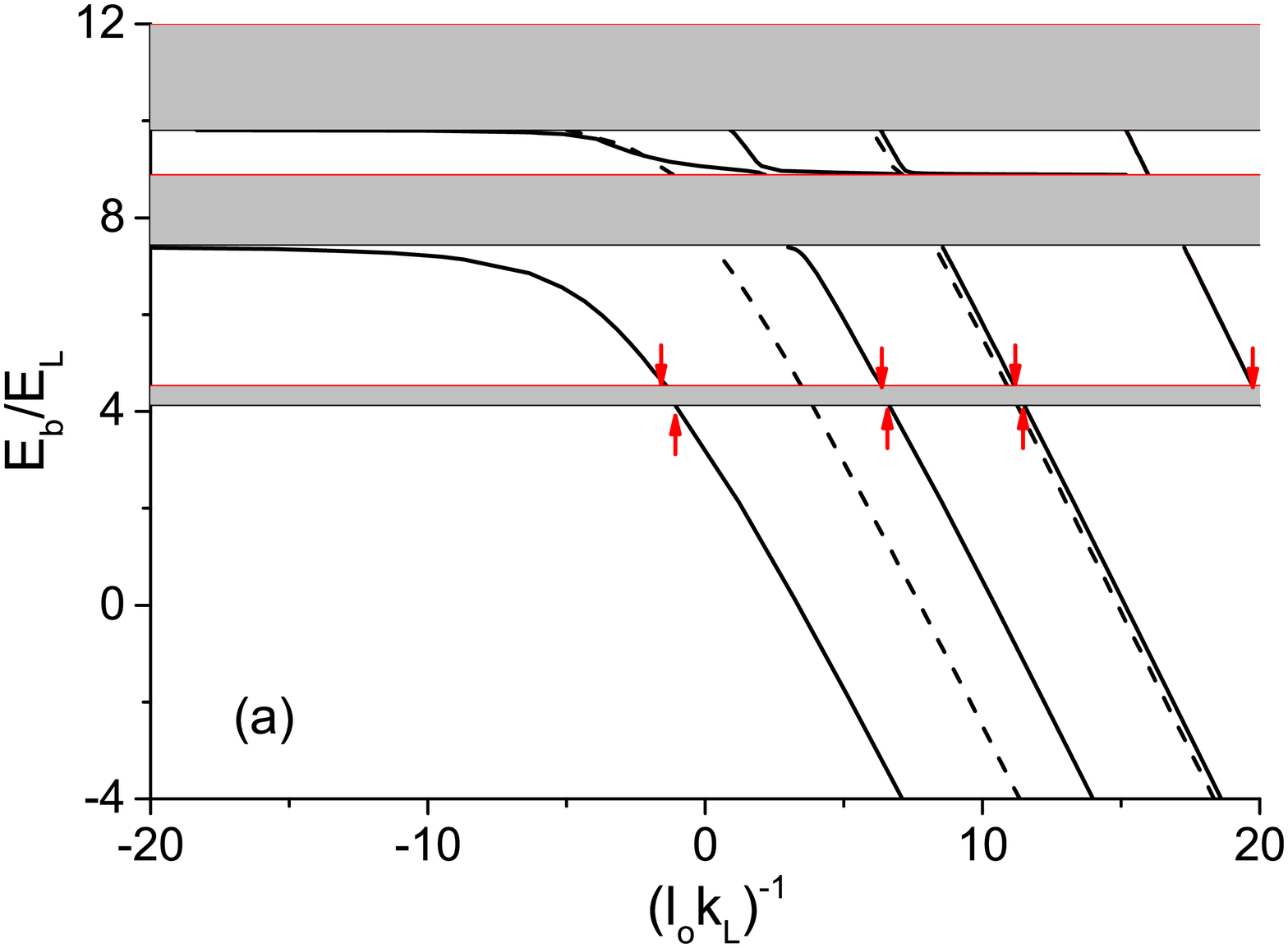} 
\includegraphics[height=4cm]{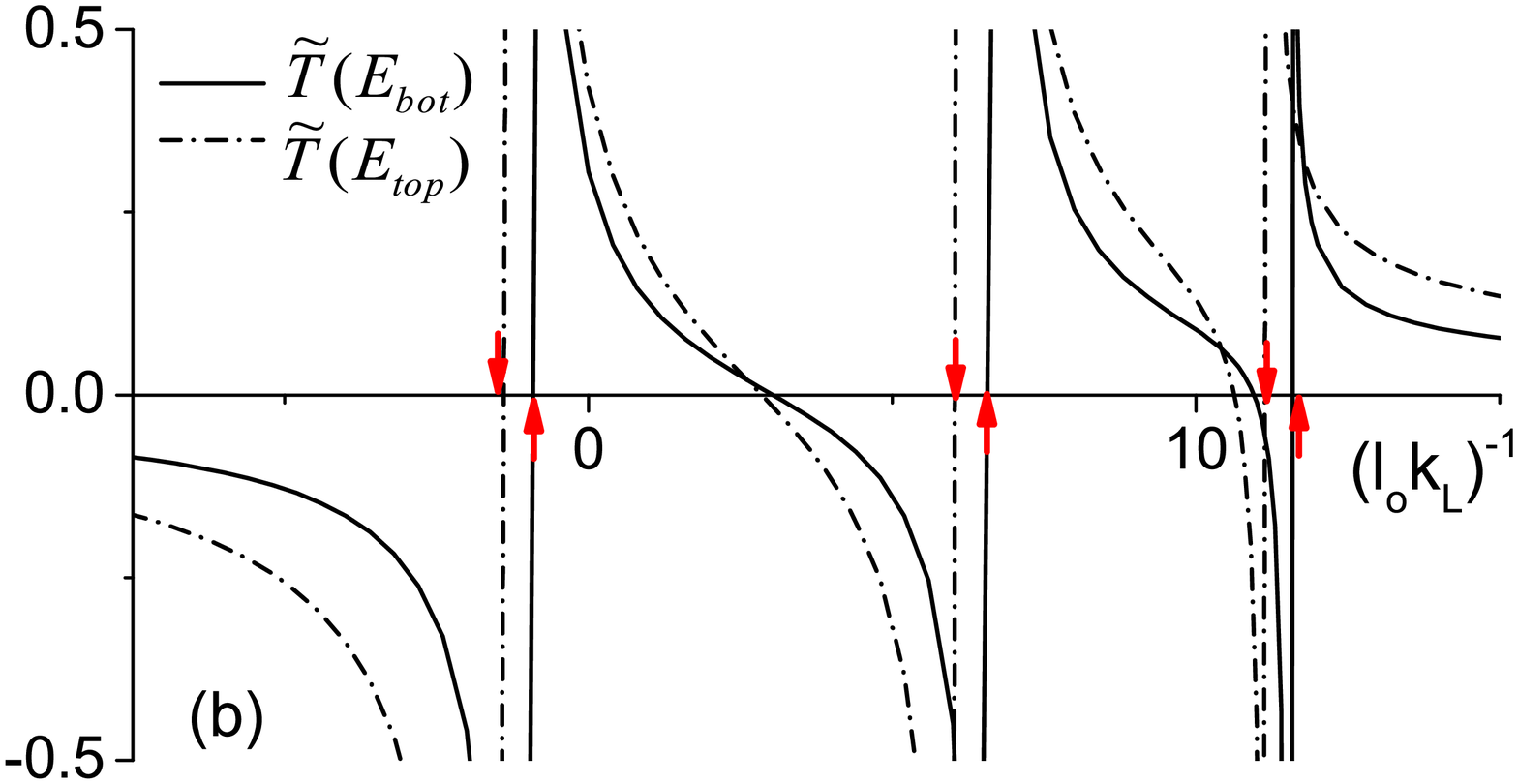} 
\caption{(Color online). (a) Bound state energy $E_b$ (in unit of $E_L$) and (b) the scaled T-matrix $\tilde{T}$ (in unit of $(mLk_L)^{-1}$) for Bloch-wave scattering near the lowest band bottom (solid lines) and top (dashed-dot) as functions of $(l_ok_L)^{-1}$. 
%Here $\tilde{T}=T/\Delta k^2$, with  $\Delta_k$ the small momentum deviated from the band bottom and top (see text).  
Here $v=6$, $r_o=0.5a_L$. In (a), the solid (dashed) lines correspond to even-parity  (odd-parity) center-of-mass motions(see text). The upward (downward) arrows in (a,b) mark the emergence of attractive (repulsive) bound states near the lowest band, where $\tilde{T}(E_{bot})$ ($\tilde{T}(E_{top})$) go across resonances.  } \label{fig1}
\end{figure}

{\it Bloch-wave resonance.} The bound states emergent at the bottom and top of each continuum band in Fig.1a imply the scattering resonances of Bloch waves at corresponding energy. %the bottom or top of each continuum band, where the bound states start to emerge. 
%To see this more clearly, we study the scattering amplitudes of two Bloch states at the bottom and the top of the lowest band, respectively denoted by $T(E_{bot})$ and $T(E_{top})$. 
In general, for any two-body scattering state $\Psi(E)=|n,k_0; n',-k_0\rangle$ ($E$ is the total energy), the on-shell scattering matrix $T(E)$ can be obtained by summing up all virtual scatterings involving the dimer and two-fermion intermediate states\cite{Cui3}. The resulting $T(E)$ can be expressed\cite{Buchler} by introducing the eigenvectors $R_N^{\alpha}$ and eigenvalues $\chi_{\alpha}$ for M-matrix (Eq.\ref{M}), which gives  
\begin{equation}
T(E)=2 \sum_{\alpha} \frac{|\sum_{N} R_N^{\alpha} c^{N}_{nn'k_0}|^2}{\frac{m}{2l_o}-\chi_{\alpha}}. \label{T_matrix}
\end{equation}
Once $m/(2l_o)$ matches one of the eigenvalues $\chi_{\alpha}$, $T(E)$ will go through a resonance, with the width determined by the nominator of above equation. %
%In combination with Eq.\ref{RG}, we see that each resonance is indeed associated with a bound state at energy $E$.

In Fig.1b, we plot the scaled T-matrix, $\tilde{T}=T/\Delta k^2$, for Bloch states scattering near the lowest-band bottom ($\Delta k=k_0\rightarrow 0$) and the top($\Delta k=k_L-k_0\rightarrow 0$) at $v=6$, respectively denoted by $\tilde{T}(E_{bot})$ and $\tilde{T}(E_{top})$. Multiple resonances are shown as tuning $1/l_o$. As only the even-N dimer levels couple with the lowest-band scattering state, these resonances are associated with the labels $\alpha=0,2,4...$ in Eq.(\ref{T_matrix}).
%Due to the odd-wave interaction, we have scaled $\tilde{T}(E_{bot})$ by $k_0^2$ and $\tilde{T}(E_{top})$ by $(k_L-k_0)^2$. 
Combined with Fig.1a, one can see that the emergences of attractive (or repulsive) bound states correspond to the resonance of $\tilde{T}(E_{bot})$ from $-\infty$ to $+\infty$ (or $\tilde{T}(E_{top})$ from $+\infty$ to $-\infty$).

\begin{figure}[h]
\includegraphics[width=7.5cm]{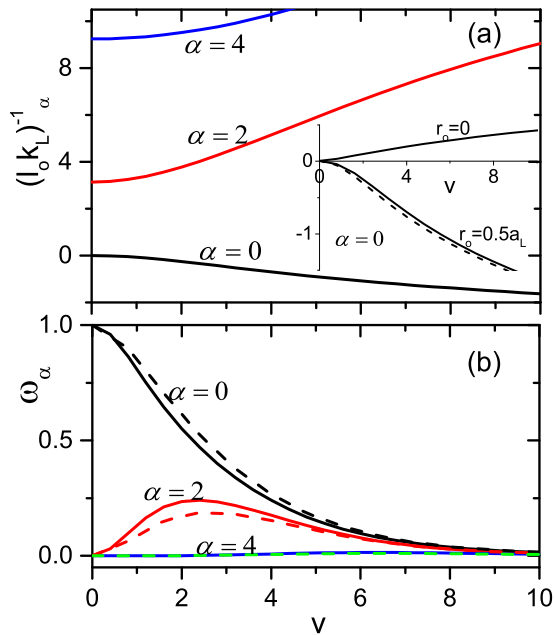} 
\caption{(Color online). Resonance locations $(l_ok_L)^{-1}_{\alpha}$ (a) and widths $\omega_{\alpha}$ (b) as functions of $v$ for the first three resonances of $\tilde{T}(E_{bot})$ in Fig.1b. Inset of (a) shows $(l_ok_L)^{-1}_{\alpha=0}$ with zero and finite ranges; the dashed line shows the prediction to finite-range case based on the decoupled-channel analysis (see text). Dashed lines in (b) shows $(c_{00;k_0}^{N=\alpha}/k_0)^2$ as $k_0\rightarrow 0$.
%, showing the same behavior as $\omega_{\alpha}$ in the main plot of (b).  
} \label{fig2}
\end{figure}

%The location and width of Bloch-wave resonances can sensitively depend on the lattice depth $v$ and the range of odd-wave interaction. 
%Given Eq.\ref{T_matrix}, the Bloch-wave resonance can be understood either as interaction-induced (by changing $l_o$) or as lattice-induced\cite{Cui} (changing $\chi_{\alpha}$. 
In Fig.2(a,b), we plot the resonance locations $(l_ok_L)^{-1}_{\alpha}$ and widths $\omega_{\alpha}=|\sum_NR_N^{\alpha}c_{00k_0}^{N}|^2/k_0^2$ for the first three resonances of $\tilde{T}(E_{bot})$ as a function of $v$. 
% the corresponding widths in the $(1/l_o, V_0)$ plane. 
We can see that as $v$ increases from zero, the first resonance ($\alpha=0$) moves from $1/l_o=0$ (free-space resonance) to  $1/l_o<0$ side (weak coupling), with decreasing resonance width; while the rest ones ($\alpha=2,4$) move to $1/l_o>0$ side (strong coupling), with the widths initially increasing and then decreasing. As shown below, these results uniquely manifest the interplay between lattices and odd-wave interactions with finite range. 
%, the interplay of lattice potential and finite effective range play essential roles in above resonance features.

First, we analyze the finite range effect to the resonance locations. 
%The range $r_o$ appears in the diagonal terms of M-matrix (Eq.\ref{M}), and thus can greatly shift the resonances. 
For comparison, in the inset of Fig.2a we plot the first resonance locations $(\alpha=0)$ for both finite and zero $r_o$. Contrary to the finite $r_o$ case, with zero $r_0$ the resonance moves to $1/l_o>0$ side as increasing $v$. A qualitative understanding can be gained from the decoupled-channel assumption, i.e., by neglecting all $M_{N\neq N'}$ in Eq.\ref{M}.
% and relating each resonance to a single dimer level $N$. 
In this case the resonance is solely determined by matching $M_{NN}$ with $m/2(1/l_o+\Delta_N)$, with $\Delta_N=mr_o(E_{bot}-E_{N0})$ denoting the difference of the N-th resonance locations between zero and finite $r_o$ cases. As shown in the inset of Fig.2a, $\Delta_{N=0}(>0)$ %gives a very good approximation to 
can well approximate the real difference for the first resonance. In large $v$ limit, $(E_{bot}-E_{00})$ is roughly the zero-point energy for the relative motion of two atoms in a single well, so we expect the first resonance occur in the very weak coupling regime $1/l_o\sim -v^{1/2}r_o/a_L^2$. Note that this should be distinguished from the induced resonance in 3D lattices with arbitrarily weak s-wave interaction\cite{Zoller, Cui, Buchler}, where the enhanced on-site coupling, rather than the range effect, plays a dominated role. %, where the finite range take the dominated effect.

% in deep lattice limit, correspond to zero-point energy of relative motion of two particles.

Second, the behavior of $\omega_{\alpha}$ can also be qualitatively understood from the decoupled-channel analysis, where $R_N^{\alpha}= \delta_{N\alpha}$ and $\omega_{\alpha}=(c_{00k_0}^{\alpha}/k_0)^2$, as shown by dashed lines in Fig.2b. %In $v\ll 1$ limit, perturbation theory gives $\omega_1=1-9v^2/256$ and $c_3=v/8$, so $\omega_1$ decreases while $\omega_3$ increases with $v$; while in $v\gg 1$ limit, the tight-binding analysis gives decaying $c_1=$ and $c_3=$. 
%We show in Fig.2b that the numerical evaluations of $c$ match these asymptotic behavior quite well. 
A remarkable feature here is that all $\omega_{\alpha}$ decay with $v$ in large $v$ limit. Physically, this is because the odd-wave interaction uniquely favors fermion-fermion correlation between neighboring lattice sites, which can be greatly suppressed by the potential barrier of deep lattices. This is in sharp contrast to the s-wave interaction case, where one of resonance widths can be greatly enhanced by on-site correlations for deep lattices\cite{Buchler}. 
%In fact, we have (exact relation: \sum_\alpha \omega_{\alpha}=...)This is why all the widths $\omega_{\alpha}$ finally decrease as increasing $v$. 
Here due to the comparable widths for large $v$ (see Fig.2b, $\omega_{0,2}$ are of the same order for $v\ge 4$), one has to treat multiple Bloch-wave resonances at equal footing. %As shown in Fig.2b, when $v\ge 4$, $\omega_0$ and $\omega_2$ are already in the same order.

\begin{figure}[h]
\includegraphics[width=6cm]{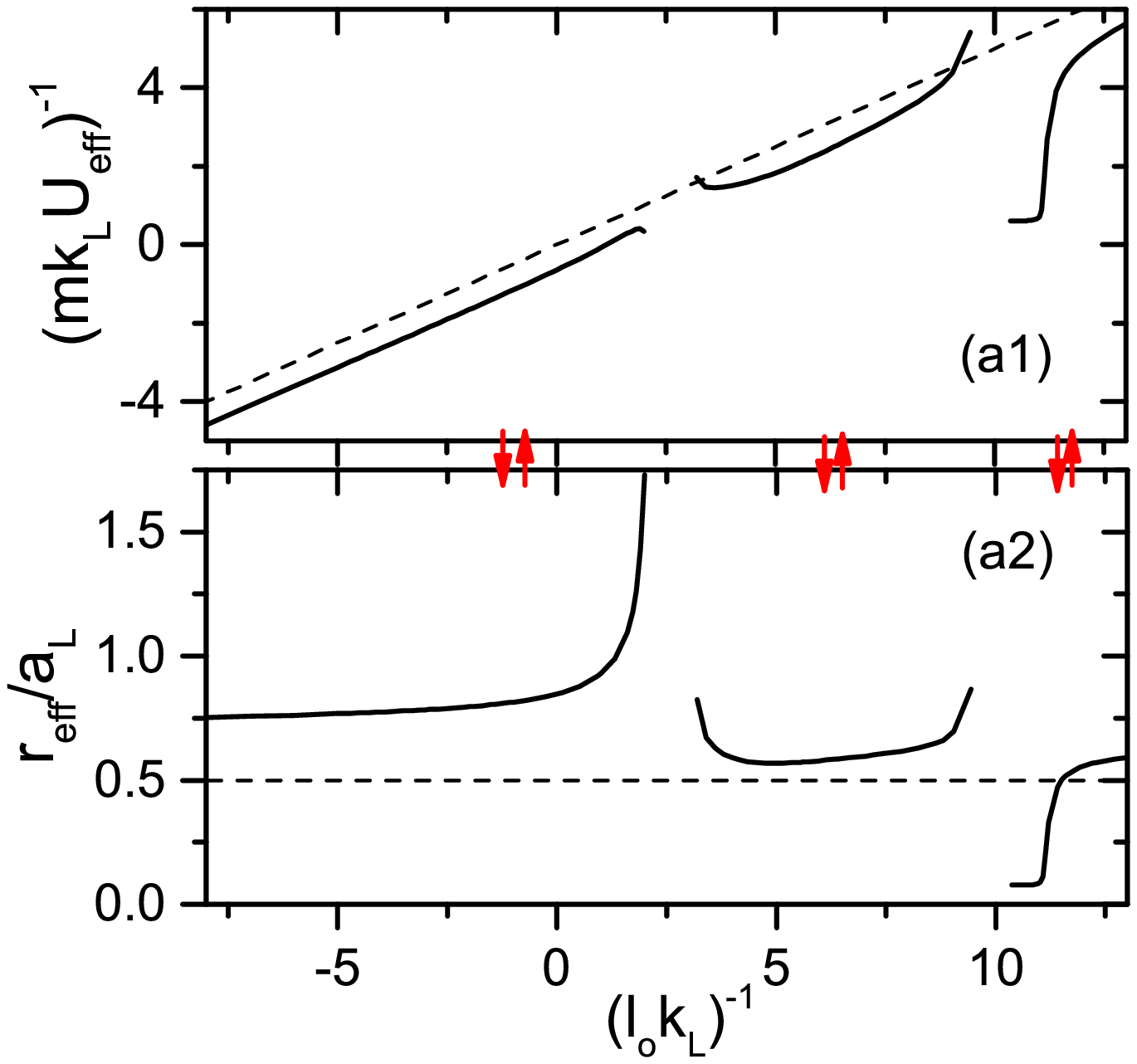} 
\includegraphics[width=8.5cm]{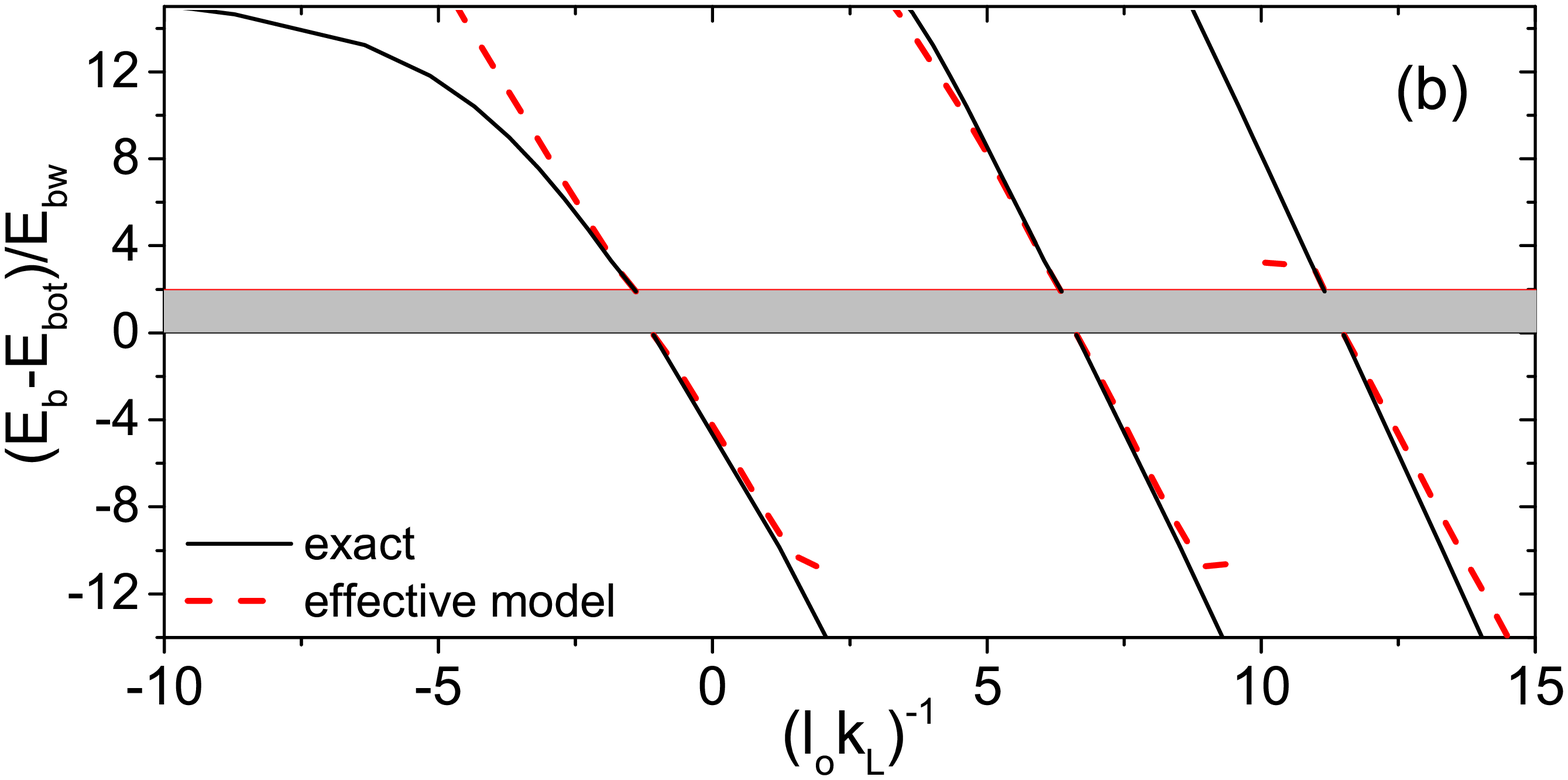} 
\caption{(Color online). (a1,a2) Effective parameters $U_{eff}^{-1}$ and $r_{eff}$ near the resonances in Fig.1b. The dashed lines show the free space results: $U_{\rm eff}=2l_o/m$ and $r_{eff}=r_o=0.5a_L$. The up- and down-ward arrows mark the resonance positions of $\tilde{T}(E_{bot})$ and $\tilde{T}(E_{top})$. (b) Bound state spectrum from exact solutions (black solid, same as Fig.1a)  and from effective model (red dashed) near the lowest band. Here energies are scaled by the single-particle band width $E_{bw}$.} \label{fig3}
\end{figure}

{\it Effective model.} Based on the two-body solutions, we can construct an effective model, $H_{eff}$, for open-channel fermions in the lowest band. 
%Given the multiple resonances that ought to be treated equally in general, we include in $H_{eff}$ all levels of closed-channel dimers. 
Namely, $H_{eff}$ corresponds to projecting the original fermion operators in Eqs.(\ref{H0},\ref{H1}) to the lowest band, while for the dimer part we keep all the bands involved considering the multiple resonances that should be treated equally in general. Accordingly, the detuning $\nu$ and coupling $g$ are then replaced by the effective ones $\nu_{eff}$ and $g_{eff}$. These two parameters resulted in an effective interaction strength $U_{eff}$ and a finite range $r_{eff}$: 
\begin{equation}
U_{eff}=-2\frac{g_{eff}^2}{\nu_{eff}};\ \ \ \ \ r_{eff}=-\frac{1}{m^2g_{eff}^2}.
\end{equation}
As $U_{eff}$ and $r_{eff}$ have encapsulated all contributions from higher-band scatterings, they can be seen as {\it regularized} interaction parameters for the lowest-band fermions.

We determine $U_{eff}$ and $r_{eff}$ by matching the T-matrix from effective model ($T_{eff}$) with the exact values ($T$ in Eq.\ref{T_matrix}) for Bloch-wave scattering at the bottom and the top of the lowest band. Specifically, $T_{eff}$ reads
\begin{equation}
T_{eff}(E)=2 \sum_{\alpha} \frac{|\sum_{N} \bar{R}_N^{\alpha} c^{N}_{00;k}|^2}{\frac{1}{U_{eff}}-\bar{\chi}_{\alpha}}, \label{T_matrix_2}
\end{equation}
where $\bar{R}_N^{\alpha}$ and $\bar{\chi}_{\alpha}$ are respectively the eigenvector and eigen-value of the matrix $\bar{M}(E)$ with elements:
\begin{eqnarray}
\bar{M}_{NN'}&=&-\frac{m^2r_{eff}}{2}(E-E_{NK}) \delta_{NN'}+\frac{1}{L}\sum_{k} \frac{{c_{00k}^{N}}^*c_{00k}^{N'} }{E-\epsilon_{0k}-\epsilon_{0,-k}}. \nonumber 
%\label{M2}
\end{eqnarray}
Thus Eq.\ref{T_matrix_2} can also predict multiple scattering resonances. 
% of Bloch waves will occur whenever $1/U_{eff}$ matches one of $\tilde{\chi}_{\alpha}$. By requiring that Eq.(\ref{T_matrix_2}) reproduce the original T-matrix for Bloch wave scattering at the top and the bottom of the lowest band
%\begin{equation}
%T(E)=\tilde{T}(E)\ \ \ \ \ (E=E_{top}, \ E_{bot})
%\end{equation}
%we can fix the effective parameters $U_{eff}$ and $r_{eff}$. To be more concrete, 
Since both %$T$ (Eq.\ref{T_matrix}) and $T_{eff}$ (Eq.\ref{T_matrix_2}) 
$T$ and $T_{eff}$ are multi-value functions of interaction parameters, to ensure a one-to-one mapping between $1/l_o$ and $1/U_{eff}$, we require $T$ and $T_{eff}$ are near the same order of resonance ($\alpha$), i.e., dominated by the same dimer level. Such procedure leads to the solution of $U_{eff}$ and $r_{eff}$ as shown in Fig.3(a1,a2), which reproduce both $\tilde{T}(E_{bot})$ and $\tilde{T}(E_{top})$ in Fig.1b. We can see that both $U_{eff}$ and $r_{eff}$ differ from the free-space results $2l_o/m$\cite{Cui2} and $0.5a_L$ (dashed lines), due to the renormalized high-band contributions. We have not determined $U_{eff}$ and $r_{eff}$ far from resonances, as in this regime the resonance order ($\alpha$) is obscure to identify. %which regime it is unclear how to identify the resonance order.
% of resonance order makes  the order of resonances making the the multi-value of T-matrix leads to multiple solutions of the parameters and the physics there is not clear the situation quite complicated.

We test the validity of $H_{eff}$ by calculating the bound state energy $E_b$, which is determined by the divergence of $T_{eff}$ in Eq.\ref{T_matrix_2} at $E=E_b$.
%  matching $1/U_{eff}$ with an eigen-energy  of $\bar{M}(E_b)$-matrix (i.e., when $T_{eff}\rightarrow\infty$ in Eq.\ref{T_matrix_2}). 
%\begin{equation}
%{\rm Det}\left( \frac{1}{U_{eff}}\delta_{NN'}-\bar{M}_{NN'}(E_b) \right)=0. 
%\end{equation}
As shown in Fig.3b, $E_b$ can well reproduce the exact solutions for attractive and repulsive bound states near Bloch-wave resonances, even within an energy range up to $5-10$ times the single-particle band width for the first two resonances. Meanwhile, we note that the effective model only works in a very narrow window for the bound states near the third resonance.  This can be attributed to its extremely narrow width $\omega_4\sim0.014\ll \omega_{0,2}$ (see Fig.2b). Accordingly, outside the resonance regime the associated bound states have much less weight in the lowest band,  and  the effective model  desired for lowest-band fermions fails to work there. 
%, will be sensitive to fermion scatterings in higher-band. As a result, the lowest-band effective model cannot capture such sensitivity and fails to describe the correct bound state solutions. We expect that the effective model will work well for low-energy physics near resonances with reasonably large width.

To this end, we have confirmed the validity of $H_{eff}$ in predicting both the scattering amplitudes and the bound states above/below the lowest-band near the Bloch-wave resonances.  Our scheme here to determine the effective parameters in $H_{eff}$ is distinct from previous studies on s-wave interacting fermions\cite{Stecher, Carr, Duan} .
We have also checked that the single-channel model, i.e., without including closed-channel dimers\cite{Buchler}, is unable to predict the correct bound states in this case. %We remark here that due to the large finite range ($r_o$ of order of $a_L$), it is important to construct the effective $H$ also in terms of two channels. We have tried various single channel models constructed by only fermion operators, with or without effective range, which are found to hardly reproduce the correct repulsive/attractive bound states simultaneously  even very close to the lowest band. 

Finally, we convert $H_{eff}$ to lattice model by expanding field operators in terms of Wannier functions, $\psi(x)=\sum_i \psi_i \omega_0(x-R_i)$; $d(x)=\sum_{N,i} d_{N,i} W_N(x-R_i)$, giving
\begin{eqnarray}
H_{eff}&=&-t_f\sum_{<i,j>}(\psi_i^{\dag}\psi_j+h.c.)-\sum_{<i,j>;N}t_d^{(N)} (d_{N,i}^{\dag}d_{N,j} +h.c.)\nonumber\\
&&+\sum_i \epsilon_f\psi_i^{\dag}\psi_i+\sum_i (\epsilon_d^{(N)}+\nu_{eff})d_{N,i}^{\dag}d_{N,i}\nonumber\\
&&+g_{eff}\sum_{N,i,\delta_1,\delta_2} \left(c_{\delta_1,\delta_2}^{(N)} d_{N,i+\delta_1}^{\dag}\psi_i\psi_{i+\delta_2} +h.c.\right), \label{H_lattice}
\end{eqnarray}
where $t_f$ and $\epsilon_f$ ($t_d^{(N)}$ and $\epsilon_d^{(N)}$) are the nearest-neighbor hopping and on-site potential of fermions (dimers at level $N$); the coupling  $c_{\delta_1,\delta_2}^{(N)}=-i\int dx W_N^*(x-\delta_1a_L) \left( \omega_0'(x)\omega_0(x-\delta_2a_L) - \omega_0(x)\omega_0'(x-\delta_2a_L) \right)$. Fixing $\delta_2=1$, we show in Fig.4a that $|c_{\delta_1,\delta_2}^{(N)}|$ is the largest when $\delta_1=0$ or $1$. We have checked that for a general $\delta_2$, $|c_{\delta_1,\delta_2}^{(N)}|$ is the largest when $\delta_1=[\delta_2/2]$ or $[(\delta_2+1)/2]$, i.e., when the dimer sits in the center of two fermions to optimize the atom-dimer coupling. In Fig.4b, we plot the largest $|c_{\delta_1,\delta_2}^{(N)}|$ as a function of $\delta_2$, and find it gradually decreases as $\delta_2$ increases from $1$. Thus to capture the most dominated atom-dimer coupling in deep lattices, we can choose the nearest-neighbor fermions ($\delta_2=1$) and the dimers sitting with either one of them  ($\delta_1=0,1$). 

\begin{figure}[t]
\includegraphics[width=8.5cm]{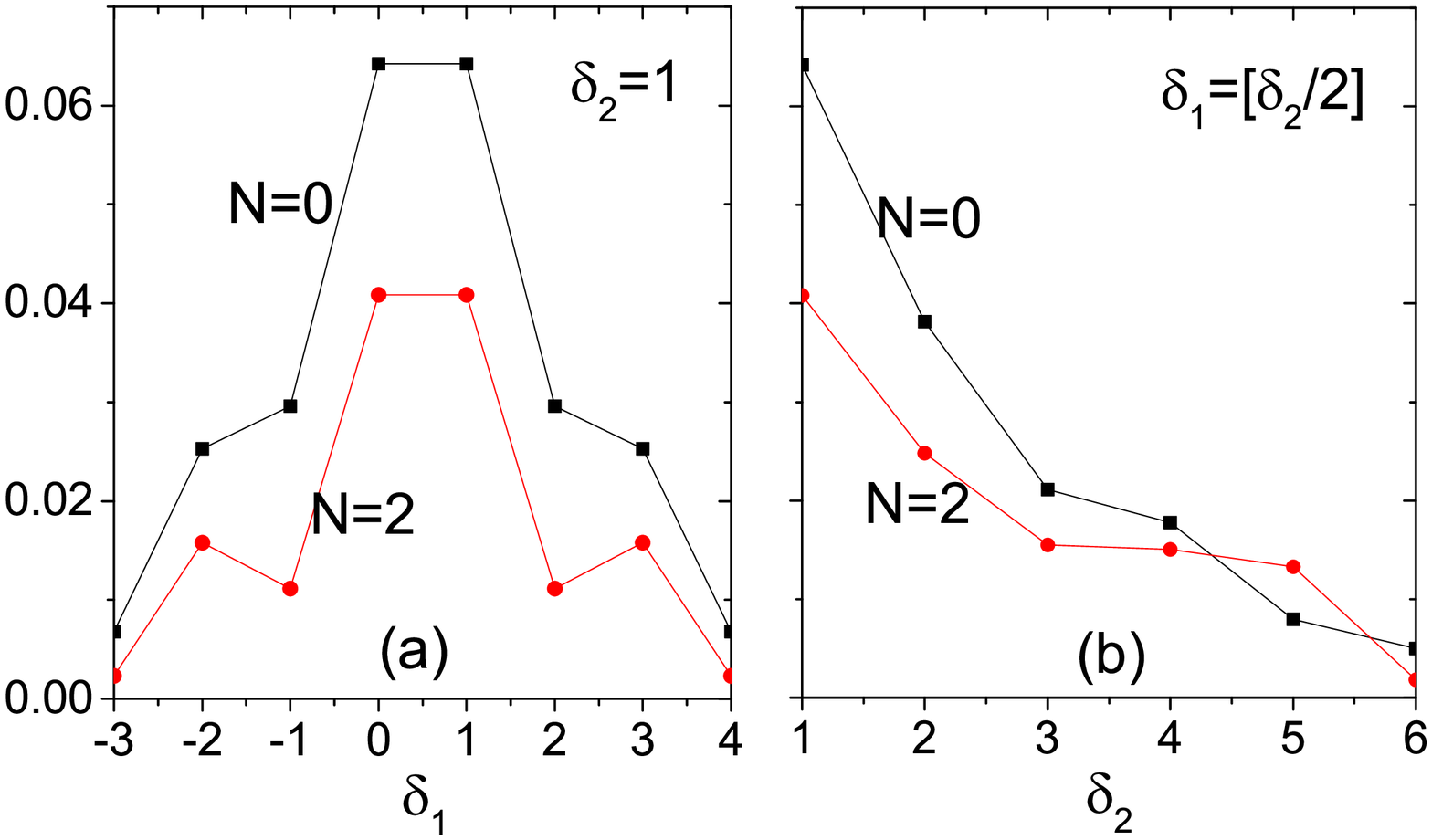} 
\caption{(Color online). Lattice parameter $|c_{\delta_1,\delta_2}^{(N)}|$ (in unit of $a_L^{-3/2}$) for $N=0,2$ at $v=6$. In (a) $\delta_2=1$; in (b) $\delta_1=[\delta_2/2]$.} \label{fig.4}
\end{figure}

Compare the lattice Hamiltonian (Eq.\ref{H_lattice}) with the Kitaev chain model\cite{Kitaev}, 
%the lattice Hamiltonian (Eq.\ref{H_lattice}) additionally includes a multiple of dimer levels. 
%These dimers together with the coupling to fermions play the role of neighbor-site pairing in the Kitaev model. Given the additional inner structure of dimers at different levels and the highly tunable interaction strength in (\ref{H_lattice}), the Majorana fermions, if exist, are expected to show a richer property in the current system.
one can expect that if the dimers condense, they will play the role of pairing mean-field in the Kitaev model and thus reproduce the Majorana physics. Meanwhile, the existence of quantum fluctuations, the resonant scatterings, and the multi-level structure of dimers in the current system are all beyond the Kitaev model. Their interplay will promisingly result in a richer many-body property in such atomic system, which is to be explored in future.
%, in this system many of other factors, including there could be strong quantum fluctuations near resonances, and the athe interplay of quantum flunctuations, the resonant coupling, and the inner structure of dimers will promisingly result in a richer property in the current system. 

%it is interesting to study how the Majorana fermions, if exist, behave in the current system. %The problem will be left for future study.
 
%\begin{equation}
%c_{l,i,j}^{(N)}=-i\int dR \bar{\omega}_N(R-\delta_1a_L) \left( \omega_1'(R)\omega_1(R-\delta_2a_L) - \omega_1(R)\omega_1'(R-\delta_2a_L) \right)
%\end{equation}

{\it Summary. } In summary, we have addressed the two-body effective scattering and bound states for 1D Fermi gases in optical lattices across odd-wave resonances. The multiple Bloch-wave resonances with comparable widths, the associated attractive/repulsive bound states, and the effect of finite interaction range can be detected in current cold atoms experiments. In addition,
%  exhibiting similar widths in deep lattice limit, in contrast to the s-wave interaction case. In addition, the large odd-wave interaction range (of order of lattice spacing) is found to greatly shift the resonance positions. These results can be directly tested in current cold atoms experiments. Finally, 
we have constructed an effective low-energy model, which successfully describes both the scattering amplitudes and bound states near the lowest band. As an analog of the Kitaev chain to host Majorana fermions\cite{Kitaev}, the effective model sets the basis for future exploring topological quantum states in realistic 1D cold atomic systems.

% studies include two-body scattering properties with finite momentum and the physics of Majorana fermions based on the effective model.
%These results set the foundation for exploring interesting many-body physics in this system in future. 

%{\it Final remarks.} Our results can be tested in the strongly odd-wave interacting quasi-1D systems with much suppressed atom loss compared to 3D case near p-wave resonances\cite{Toronto, K40, K40_2, Li6_1,Li6_2}. Such suppression is indicated by the absence of centrifugal barrier for 1D collision and thus the spatially much extended bound state (see Fig.2b), which is less likely to decay into deep molecules and cause three-body losses. Moreover, the shifted resonance in quasi-1D\cite{CIR_p_1,CIR_p_2,CIR_p_3} also helps to avoid severe losses near the odd-wave resonance, whose location in magnetic field can be far from that of a 3D p-wave resonance.

{\it Acknowledgment.}  The work is supported by the National Natural Science Foundation of China (No.11626436, No.11374177, No. 11421092, No. 11534014), the National Key Research and Development Program of China (2016YFA0300603), and the National Science Foundation under Grant No. NSF PHY11-25915. The author would like to thank the hospitality and support of the Kavli Institute for Theoretical Physics in Santa Barbara during the program "Universality in Few-Body Systems" in the winter of 2016, where this manuscript was partly finished.

\end{document}